\documentclass[journal=jpcafh,manuscript=article,layout=twocolumn]{achemso}
\setkeys{acs}{articletitle = true}
\usepackage[percent]{overpic}
\usepackage{color}
\usepackage{gensymb}
\usepackage[table]{xcolor}
\usepackage{hyperref}
\usepackage{xcite}
\usepackage{xr}
\usepackage{amsmath}
\usepackage{amssymb}
\usepackage{lipsum}

\usepackage[version=3]{mhchem} 

\usepackage{bbold}


\author{\small Braden M. Weight}
\affiliation{\small Theoretical Division, Center for Integrated Nanotechnologies, Los Alamos National Laboratory, Los Alamos, NM 87545, U.S.A.}
\email{braden.m.weight@lanl.gov}

\author{\small Aaron Forde}
\affiliation{\small Theoretical Division, Center for Integrated Nanotechnologies, Los Alamos National Laboratory, Los Alamos, NM 87545, U.S.A.}

\author{\small Sergei Tretiak}
\affiliation{\small Theoretical Division, Center for Integrated Nanotechnologies, Los Alamos National Laboratory, Los Alamos, NM 87545, U.S.A.}
\email{serg@lanl.gov}

\title{\large A Visual Understanding of Circular Dichroism Spectroscopy: The Case of Interacting Chromophores}

\begin{document}

{\footnotesize
\begin{abstract}
Exciton coupling shapes the photophysical and chiroptical properties of molecular aggregates, but a microscopic, real-space understanding of how intermolecular interactions generate circular dichroism (CD) remains incomplete. Here, we computationally investigate the microscopic origins of CD in coupled molecular dimers to separate intermolecular (topological) from intramolecular (intrinsic) chirality. A combination of time-dependent density functional theory and transition chiral tensor (TCT) analysis provides a quantitative, visually intuitive framework showing how electronic coupling strength, intermolecular separation, and relative angular orientation govern chiroptical response. We find that pairs of individually achiral chromophores exhibit robust, topological chiroptical signals which give rise to the conventional Cotton effect, a standard interpretation of derivative features in circular dichroism spectra. We next establish that intrinsically chiral chromophores, on the other hand, disrupt the Cotton effect due to the competition between intrinsic chirality and topological chirality. 
This work establishes a foundation for interpreting exciton-coupled CD spectra in molecular aggregates, supramolecular assemblies, and nanostructures using the TCT approach, which offers an attractive tool for analyzing chiroptical responses in complex multi-chromophore systems relevant to emerging quantum technologies.
\end{abstract}
}

\section{Introduction}
Circular dichroism (CD) spectroscopy has emerged as an indispensable tool for probing chiral molecular structures and their electronic properties. The differential absorption of left- and right-circularly polarized light provides unique insights into molecular stereochemistry, protein secondary structure, and the organization of supramolecular assemblies.\cite{scholes_molecular_2023,nhon_critical_2025,zelevinsky_physical_2023,kagan_colloidal_2021,alfieri_nanomaterials_2023,head-marsden_quantum_2021,von_kugelgen_chemical_2019,wu_foundations_2024,wasielewski_exploiting_2020,liu_2d_2019,scholes_quantum_2025,singh_preparing_2022} Recently, the exciting field of chiral induced spin-selectivity (CISS) has shown improved functionality for chiral materials which can selectively transmit electrons with a preferred spin orientation.\cite{bloom_chiral_2024,firouzeh_chirality-induced_2024,naaman_chiral_2020} While CD spectroscopy is widely applied across chemistry, biology, and materials science, a molecular-level understanding of how electronic coupling\cite{hestand_molecular_2017,hestand_expanded_2018,kasha_exciton_1965,weight_coupling_2021} between chromophores influences chiroptical properties remains an active area of investigation.

In molecular aggregates and complex nanostructures, individual chromophore units interact through electronic coupling (i.e., a combination of Coulomb and exchange interactions), leading to delocalized (or hybridized) excited excitonic states that can dramatically alter optical properties compared to isolated molecules.\cite{hestand_molecular_2017,hestand_expanded_2018,kasha_exciton_1965} This phenomenon is particularly significant in biological light-harvesting complexes, where precisely arranged pigments exploit electronic coupling for efficient energy transfer.
Similarly, synthetic molecular aggregates and nanostructures exhibit rich spectroscopic behavior arising from collective electronic excitations\cite{weight_coupling_2021,zheng_guanine-specific_2022,zheng_photoluminescence_2021,lin_dna-guided_2022,forde_induced_2022} which are expected to have interesting chiroptical behaviors yet to be fully understood.\cite{tiffany_inducing_2025,tiffany_collective_nodate,forde_induced_2022,forde_influence_2024,forde_molecular_2023}

The theoretical framework for understanding exciton coupling effects to describe the chiroptical properties of interacting chromophores has been established,\cite{berova_exciton_1999} which builds upon previous works exploring interacting excitons,\cite{kasha_exciton_1965} going beyond the simple Kasha model and including Dexter-like effects.\cite{hestand_molecular_2017,hestand_expanded_2018} The interacting exciton model contains two terms (Eq.~\ref{EQ:R_DIMER}), one of which represents the usual $\propto \sum_{ij}\vec{d}_i\cdot\vec{m}_j$ electric dipole - magnetic dipole interference while the second is gauge-correcting (or origin-correcting) term $\propto \omega \sum_{ij} (\vec{R}_{i} - \vec{R}_{j}) (\vec{d}_i \times \vec{d}_j)$ with $\vec{R}_i$ the center-of-mass positions of the point dipoles. See \textbf{Method} for additional details. Importantly, for achiral chromophores, the usual interference term disappears, but the gauge term survives. For this reason, the usual interference term is usually neglected when considering extended systems. For chiral chromophores, both terms survive and can either enhance or reduce the overall CD response. In this case, both terms must be kept. The competition between these two terms will be immediately evident during our discussion below.

These model approaches, while providing valuable intuition, often require quantum mechanical parameterization for accurate predictions, especially when dealing with strongly coupled systems or spatially overlapping chromophores where Dexter interactions are present.\cite{hestand_molecular_2017,hestand_expanded_2018} Advances in quantum chemistry have enabled detailed quantum mechanical investigations of chiroptical properties, such as time-dependent density functional theory (TDDFT) for both small\cite{freixas_chiral_2025,khanna_deconstructing_2025} and large systems.\cite{weight_visual_2026,tiffany_inducing_2025,tiffany_collective_nodate,forde_induced_2022,forde_molecular_2023} Such \textit{ab initio} treatment captures both short-range exchange interactions and long-range Coulombic coupling, providing a nearly complete quantitative description of exciton interactions. Using these methods, a recently developed analysis tool for visualizing chiroptical response in electronic excitations has enhanced our ability to interpret complex chiroptical responses in terms of molecular structure and electronic coupling.\cite{weight_visual_2026,weight_influence_2026,freixas_chiral_2025} This approach, as we will show, is ideal for analyzing systems in the coupled exciton picture due to its ability to decompose the chiroptical response into real-space or atom-centered detail, thus disentangling the intra- (\textit{i.e.}, dipole interference) and inter- (\textit{i.e.}, gauge) molecular effects.

Furthermore, despite recent advances, the effects of key excitonic coupling parameters on CD spectra remain insufficiently understood. These include intermolecular distance $L_{z}$, chromophore orientation $\theta$, and most importantly, intrinsic chromophore chirality. Their interplay across different coupling regimes has yet to be systematically explored. Such understanding is essential for designing new chiroptical materials, interpreting experimental spectra, and optimization of photonic devices that exploit chiral light-matter interactions.

\begin{figure}[t!]
    \centering
    \includegraphics[width=0.9\linewidth]{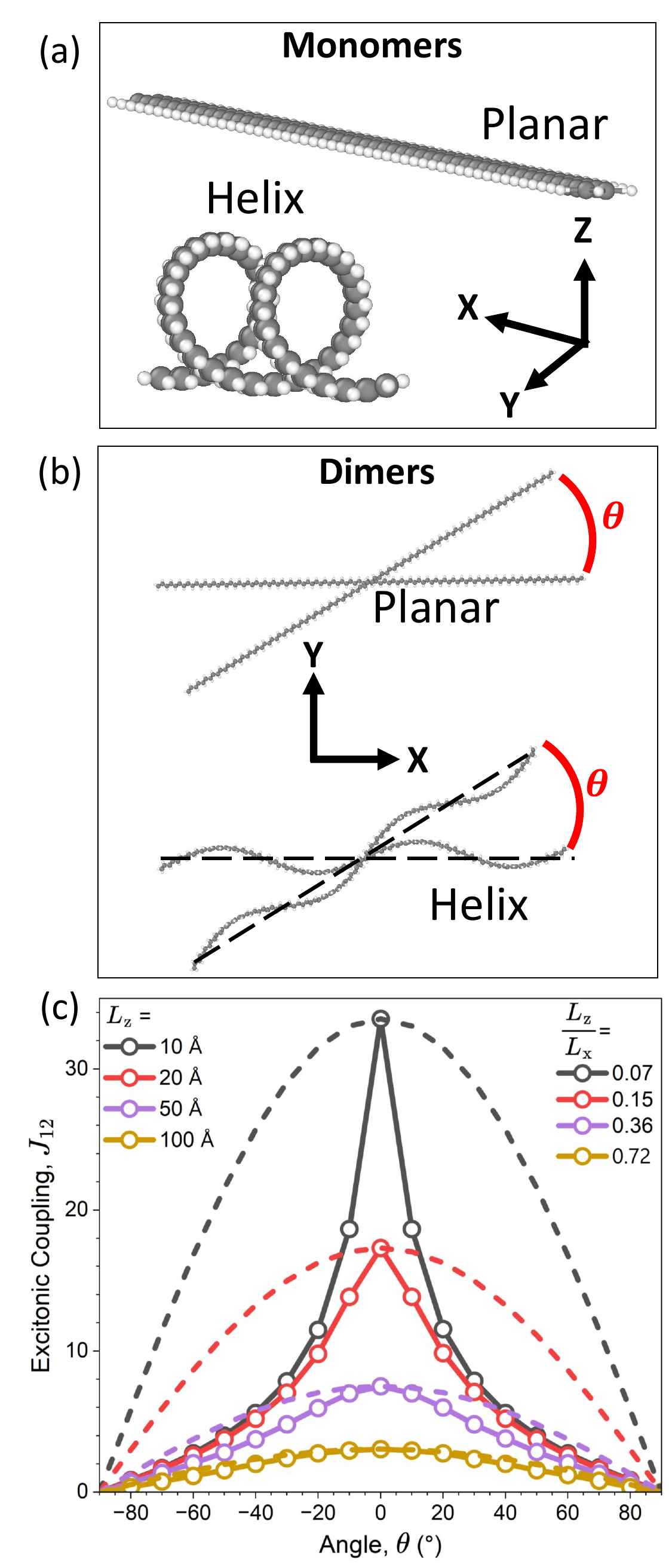}
    \caption{\footnotesize
    (a) Planar and Helix polyacetylene chromophore geometries. (b) Planar and Helix polyacetylene dimers. Each chromophore is shifted up or down by $L_z/2$ and where the shifted-up chromophore is rotated about the $z$-axis by $\theta$. (c) Exciton coupling $J_{12} = (E_+ - E_-)/2$ extracted from TD-DFT energy splitting as function of dipole-dipole angle for a variety of separation lengths $L_{z}$ = 10, 20, 50, and 100 $\AA$ for a pair of planar chromophores. The dashed lines correspond to a cosine model $J_{12} \propto \cos(\theta)$ scaled by the $\theta = 0\degree$ coupling from TDDFT. The chromophore length to dimer separation length ratio is shown as a second legend.
    }
    \label{FIG:geom_scheme}
\end{figure}

\section{Results and Discussion}

In this work, we consider the prototypical \textit{cis}-polyacetylene molecule, with composition H$_2$C(CH)$_{n-2}$CH$_{2}$ where $n = 100$, as a model molecular system. Figure~\ref{FIG:geom_scheme}a shows the polyacetylene in two geometries: ``Planar'' and ``Helix''. We refer to the individual polyacetylene molecules as chromophores. This, and similar, chemical systems have been used recently in the exploration of chiroptical response.\cite{forde_helically_2026,weight_visual_2026} Here, we extend this system to a coupled molecular case. Figure~\ref{FIG:geom_scheme}b shows the two dimers explored in this work. The chromophores are vertically separated along the $z$-axis and positioned symmetrically about the origin. The top molecule is rotated by an angle $\theta$. The dimer composed of planar chromophores is a chiral system where its constituents are achiral, and the dimer composed of helical chromophores entangles two types of chirality -- that of the multi-chromophore length scale and that of the individual chromophores. Competition between types of chirality is at the forefront of designing and controlling chiroptical responses.\cite{tiffany_inducing_2025,tiffany_collective_nodate,weight_influence_2026}

The planar polyacetylene chromophore belonging to the $C_{2h}$ symmetry group has an optically active, bright transition between its ground electronic state and its first excited electronic state of $1A_g\rightarrow1B_u$ symmetry. The transition is dominated by $\pi-\pi^*$ character and thus has a strong electric transition dipole, $\vec{d} = (19.51, 1.57, 0.0)$, aligned with the main axis of the chromophore ($x$-axis). However, by symmetry, its corresponding magnetic transition dipole, $\vec{m} = (0, 0, 0)$, is zero. Thus, the chromophore does not have a chiroptical response due to vanishing rotary strength $R$, which is defined as $R\propto \vec{d} \cdot \vec{m} = |\vec{d}| |\vec{m}| \cos(\phi_{\vec{d} \cdot \vec{m}})$. However, for the case of the helical chromophore, the applied torsion generating out-of-plane  rotation breaks the mirror and inversion symmetries , activating the magnetic transition dipole (or transition orbital angular momentum). The corresponding electric and magnetic transition dipoles are $\vec{d} = (16.51, -0.77, -0.10)$ and $\vec{m} = (-6.02, -5.41, -0.71)$, respectively, and whose mutual angle is $\phi_{\vec{d} \cdot \vec{m}} \approx 45\degree$ ($\cos(\phi_{\vec{d} \cdot \vec{m}}) \approx 0.7$). Thus, the helical chromophore retains much of the character of the planar chromophore's electronic transition in that the electric transition dipole is still dominated by the primary axis of the polymer ($x$-axis) but with non-zero transition orbital angular momentum whose direction is partially aligned with the main axis.

We next turn to the coupled chromophore systems with the exciton energies $|E_\pm\rangle$ (Davydov's pair of excited states) as functions of the intermolecular distance $L_{z}$ and relative electric transition dipole orientation $\theta = \cos^{-1}\big(\frac{\vec{d}_1\cdot\vec{d}_2}{|\vec{d}_1||\vec{d_2}|}\big)$. Here, $\vec{d}_i$ refers to the electric transition dipole moment of chromophore $i$. Figure~\ref{FIG:geom_scheme}c presents the exciton coupling strength extracted from the TDDFT exciton energy splittings as $J_{12} = \frac{E_+ - E_-}{2}$ for a given intermolecular separation $L_{z}$ as functions of the orientation $\theta$. At larger separations $L_{z} \ge 50 \mathrm{\AA}$, the interaction follows a cosine function, \textit{e.g.}, $J_{12} \propto \cos(\theta)$, with a maximum at $\theta = 0\degree$, which is the maximum for all functions since the two transition dipoles are maximally aligned. For small separation lengths, $L_{z} < 50 \mathrm{\AA}$, the interaction follows an exponential function scaling the cosine, \textit{e.g.}, $J_{12} \propto e^{-\theta}\cos(\theta)$. For clarity, the right-sided legend shows the ratio of the intermolecular separation $L_{z}$ to the length of the chromophore $L_\mathrm{x}$. For large separations, $\frac{L_{z}}{L_\mathrm{x}} \approx 1$, the point-dipole approximation is valid, while for $\frac{L_{z}}{L_\mathrm{x}} << 1$, this approximation breaks down. For comparison, the dashed lines correspond to the function $\cos(\theta)$ scaled by the $\theta = 0\degree$ TDDFT coupling. Overall, these choices of intermolecular lengths provides a range of couplings from weakly interacting point-dipole-like densities to strongly interacting densities.

\begin{figure}[t!]
    \centering
    \includegraphics[width=0.96\linewidth]{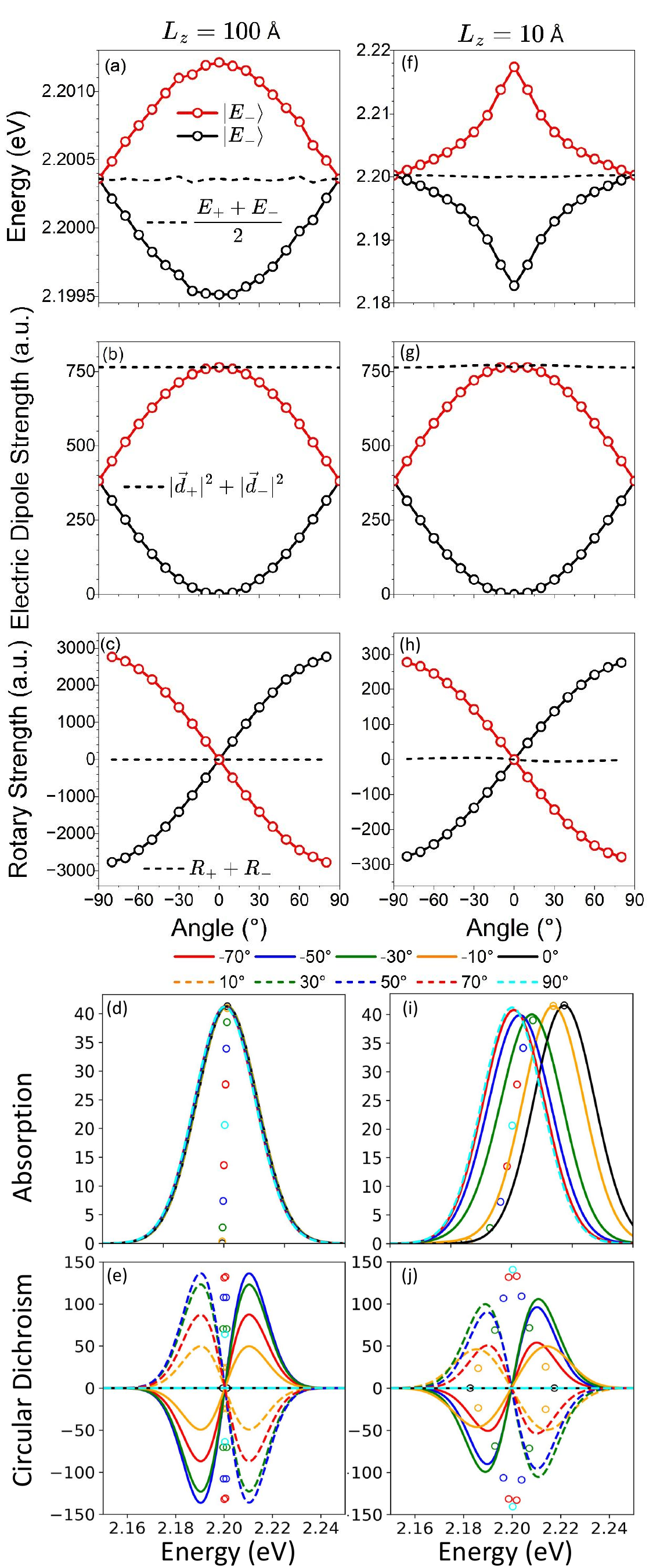}
    \caption{\footnotesize
    Exciton (a,f) energies $E_\pm$, (b,g) dipole strengths $|\vec{d}_\pm|^2$, (c,h) rotary strengths $R_\pm$, (d,i) absorption spectra and (e,j) CD spectra for a pair of \textit{achiral} chromophores separated by (a-d) $L_z = 100~\mathrm{\AA}$ (weakly coupled) and (e-h) $L_z = 10~\mathrm{\AA}$ (strongly coupled). The horizontal dashed black lines are conserved quantities for all coupling parameters. The spectra was generated with phenomenological Gaussian broadening $\sigma = 10$ meV. The symbols in the spectra indicate scaled oscillator and rotary strength values.
    }
    \label{FIG:achiral_E_D_R_ABS_CD_10_100}
\end{figure}

\begin{figure*}[t!]
    \centering
    \includegraphics[width=\linewidth]{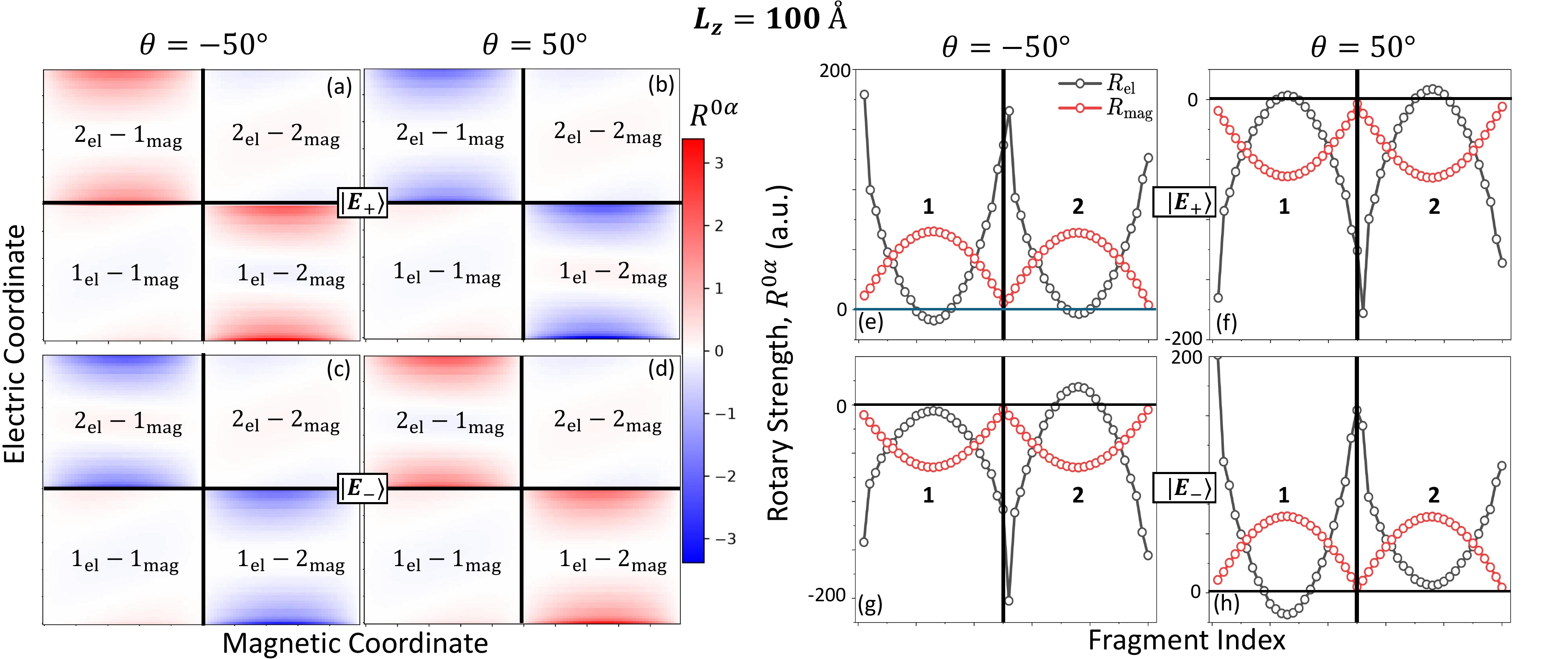}
    \caption{\footnotesize
    (a-d) Transition chiral tensors (TCTs) and (e-h) electric (black) and magnetic (red) projections of the TCTs for enantiomeric angles $\theta$ = (a,c,e,g) $-50\degree$ and (b,d,f,h) $50\degree$ for a dimer composed of two \textit{achiral} chromophores spaced at $L_{z} = 100~\mathrm{\AA}$ (weakly coupled) for both the (a,b,e,f) $|E_+\rangle$ and (c,d,g,h) $|E_-\rangle$ excitons. The atomic orbital contributions are summed into each chromophores' minimal C$_2$H$_2$ fragment as $R_{AB} = \sum_{\mu \in A} \sum_{\nu \in B} R_{\mu\nu}$ where $\{\mu,\nu\}$ are atomic orbitals and $\{A,B\}$ are the C$_2$H$_2$ fragments.
    }
    \label{FIG:achiral_dimer_TCTs_Lz_100_PROJECTIONS}
\end{figure*}

{\bf Achiral Substituents}

Figure~\ref{FIG:achiral_E_D_R_ABS_CD_10_100}a-c presents the exciton energies $E_\pm$ (panel a), electric dipole strength $|\vec{d}_\pm|^2$ (panel b), and rotary strength $R_\pm$ (panel c) as functions of the interaction angle $\theta$ for a fixed intermolecular separation length of $L_{z} = 100 \mathrm{\AA}$ (\textit{i.e.}, in the point-dipole regime, see Figure~\ref{FIG:geom_scheme}). Figure~\ref{FIG:achiral_E_D_R_ABS_CD_10_100}f-h shows the same observables for the strongly coupled case ($L_z = 10~\mathrm{\AA}$). As expected, when the chromophores are perfectly orthogonal to one another, $\theta = \pm 90\degree$, the coupling strength vanishes (see Eq.~\ref{EQ:J_EL} in \textbf{Methods}). In contrast, the coupling is maximized at $\theta = 0\degree$ when the electric transition dipole moments of the two chromophores are parallel. As expected, the splitting of the two chromophore excitons is nearly symmetric in energy (see dashed black line in Figure~\ref{FIG:achiral_E_D_R_ABS_CD_10_100}a which traces the average energy of the coupled exciton states, $(E_+ - E_-)/2 \approx \omega$), owing to a large separation in energy between the chromophore S$_1$ and S$_2$ states as well as vanishingly weak dipole interactions between $E_+$ and other S$_\mathrm{n}$ states. The effects of the energetic splitting differ due to the strong and weak coupling discussed above in Figure~\ref{FIG:geom_scheme}c

The electric dipole strength (Figure~\ref{FIG:achiral_E_D_R_ABS_CD_10_100}b) shows the sharing of the diabatic character between the two chromophore excitons. Since the chromophores are identical and have degenerate transition energies $\omega$, the coupled wavefunctions are $|E_\pm\rangle = \frac{1}{\sqrt{2}}[ |1\rangle \pm |2\rangle ]$. As noted above, there is minimal interaction with states outside the S$_1$ subspace, as shown by the constant sum of the dipole strength as a function of the interaction angle $\theta$. The dipole strengths sum to $2|\vec{d}|^2$ since $|\langle E_\pm | \vec{\hat{d}} | \mathrm{G} \rangle|^2 = \langle E_\pm | \vec{\hat{d}} | \mathrm{G} \rangle \cdot \langle \mathrm{G} | \vec{\hat{d}} | E_\pm \rangle = \sum_{ij} C_{i,\pm}C_{j,\pm} \langle i | \vec{\hat{d}} |\mathrm{G}\rangle \cdot \langle \mathrm{G} | \vec{\hat{d}} |j\rangle = \sum_{ij} C_{i,\pm}C_{j,\pm} \vec{d}_i \cdot \vec{d}_j = |\vec{d}|^2 \big(1 \pm \cos(\theta)\big)$. We used the fact that $\langle 1 | \pm \rangle = 1/\sqrt{2}$ and $\langle 2 | \pm \rangle = \pm1/\sqrt{2}$. Thus, it follows that $|\langle E_+ | \vec{\hat{d}} | \mathrm{G} \rangle|^2 + |\langle E_- | \vec{\hat{d}} | \mathrm{G} \rangle|^2 = 2|\vec{d}|^2$ at $\theta = 0\degree$ and 0 for $\theta = \pm90\degree$. The trends in dipole strength are the same between the weakly coupled ($L_z = 100~\mathrm{\AA}$) and strongly coupled ($L_z = 10~\mathrm{\AA}$) cases since the mixing of the two chromophore wavefunctions is independent of the magnitude of the coupling, even in the case where there are multipolar (\textit{i.e.}, beyond dipole) contributions to the coupling.

The behavior of the rotary strengths, shown in Figure~\ref{FIG:achiral_E_D_R_ABS_CD_10_100}c,i, is more intriguing. At maximal dimer coupling $\theta = 0\degree$, the energy splitting and electric-dipole character, and consequently the absorption intensity of the bright coupled state $|E_+\rangle$, are largest, while the rotatory strength of both states vanishes by symmetry. The rotary strength is instead maximized at the weakest coupling configurations $\theta = \pm 90\degree$, which corresponds to the left-handed ``-'' or right-handed ``+'' configurations (\textit{i.e.}, when the turns are counter-clockwise or clockwise around the z-axis). Note that each chromophore, alone, is achiral and does not give a chiral response. This is evidenced by the fact that the sum of the rotary strengths of the $|E_+\rangle$ and $|E_-\rangle$ states is zero for all configurations. 

Notably, the rotary strength (Figure~\ref{FIG:achiral_E_D_R_ABS_CD_10_100}c,h) differs, between the $L_z = 10~\mathrm{\AA}$ (strongly coupled) and the $L_z = 100~\mathrm{\AA}$ (weakly coupled) cases, by a factor equal to their separation length ratio $L_z(10~\mathrm{\AA})/L_z(100~\mathrm{\AA})$. This is due to the gauge-correcting term in Eq.~\ref{EQ:R_DIMER}, which is proportional to the intermolecular separation length, $|\vec{R}_{12}| = |\vec{R}_1 - \vec{R}_2|$. Supporting Figure~\ref{FIG:model_achiral} presents the results of a coupled exciton model for a pair of achiral chromophores which reproduces the same trends as shown in Figure~\ref{FIG:achiral_E_D_R_ABS_CD_10_100}.

Due to the difference in symmetry of the two $|E_\pm\rangle$ states, one showcases a positive chiral signal while the other shows an equal but oppositely signed signal at any given configuration. Importantly, this implies that the CD spectra will not showcase any detectable signature at $\theta = \pm 90\degree$ (nor at $\theta = 0\degree$, an achiral arrangement) due to cancellation of positive and negative degenerate-in-energy signals. However, at intermediate angles $0 < |\theta| < 90\degree$, the dimer exhibits intermediate couplings (\textit{i.e.}, the energy splitting between the between the $|E_\pm\rangle$ states) and rotary strengths, giving rise to the characteristic Cotton effect shown in Figure~\ref{FIG:achiral_E_D_R_ABS_CD_10_100}e,j, a standard interpretation of derivative features in circular dichroism spectra. While the CD spectra showcases angle-dependent trends as well as a clear separation between the two hybrid states $|E_\pm\rangle$, the absorption spectra exhibits much less information. 

Figure~\ref{FIG:achiral_E_D_R_ABS_CD_10_100}d,i show the absorption spectra for the weakly coupled (panel d) $L_z = 100~\mathrm{\AA}$ and strong coupled (panel i) $L_z = 10~\mathrm{\AA}$ cases. The weakly coupled case shows negligible variation between configurations due to the small energy splitting. Note that a phenomenological Gaussian broadening has been applied with width $\sigma = 10$ meV. Thus, for the weak coupling case, the coupling $J$ strength is less than the broadening $\sigma$. However, for the strongly coupled case, the coupling $J$ is on the same order of magnitude as the broadening $\sigma$, giving rise to the characteristic blue shift of the absorption spectra for H-aggregate like couplings. Note that this is also the reason that the intensities of the CD spectra are similar, even though the rotary strengths differ by an order of magnitude. Thus, the magnitude of static and dynamic disorder will play a large role in resolving topological chirality in experiment. \textit{Thus, the CD spectra exhibits configuration-dependent spectral signatures, whereas the absorption does not.}

Using a recently developed transition chiral tensor (TCT) analysis method,\cite{weight_visual_2026,freixas_chiral_2025,forde_helically_2026} we are able to better understand the microscopic origin of the chiroptical response by examining the response in real-space by decomposing the rotary strength into atomic orbital contributions and summing the response across sub-chromophore fragments (\textit{e.g.}, C$_2$H$_2$ units) or across the entire chromophore. See Ref.~\citenum{weight_visual_2026} for more details on the analysis method. Figure~\ref{FIG:achiral_dimer_TCTs_Lz_100_PROJECTIONS}a-d presents the TCT for the enantiomeric angles, $\theta$ = (a,c) $-50\degree$ and (b,d) $50\degree$, for both the (a,b) upper $|E_+\rangle$ and (c,d) lower $|E_-\rangle$ coupled states at fixed separation length $L_z = 100~\mathrm{\AA}$. Each panel is split into four quadrants which represent the intra- (\textit{i.e.}, 1-1 and 2-2) or inter-molecular (\textit{i.e.}, 1-2 and 2-1) electric and magnetic interference patterns. The two axes represent the (vertical) electric and (horizontal) magnetic transition dipole constituents and all of their mutual interactions, $R_{AB} \propto \vec{d}_{AB}\cdot\vec{m}_{AB}$. Each block is labeled by the chromophore-chromophore identities and their corresponding electric or magnetic contribution. Note that the rotary strength plotted in Figure~\ref{FIG:achiral_E_D_R_ABS_CD_10_100}c is recovered upon summation of the TCT as $R_\pm = \sum_{AB} R_{AB}$, \textit{i.e.}, summing over all four blocks of the tensor.\cite{weight_visual_2026}

\begin{figure*}[t!]
    \centering
    \includegraphics[width=\linewidth]{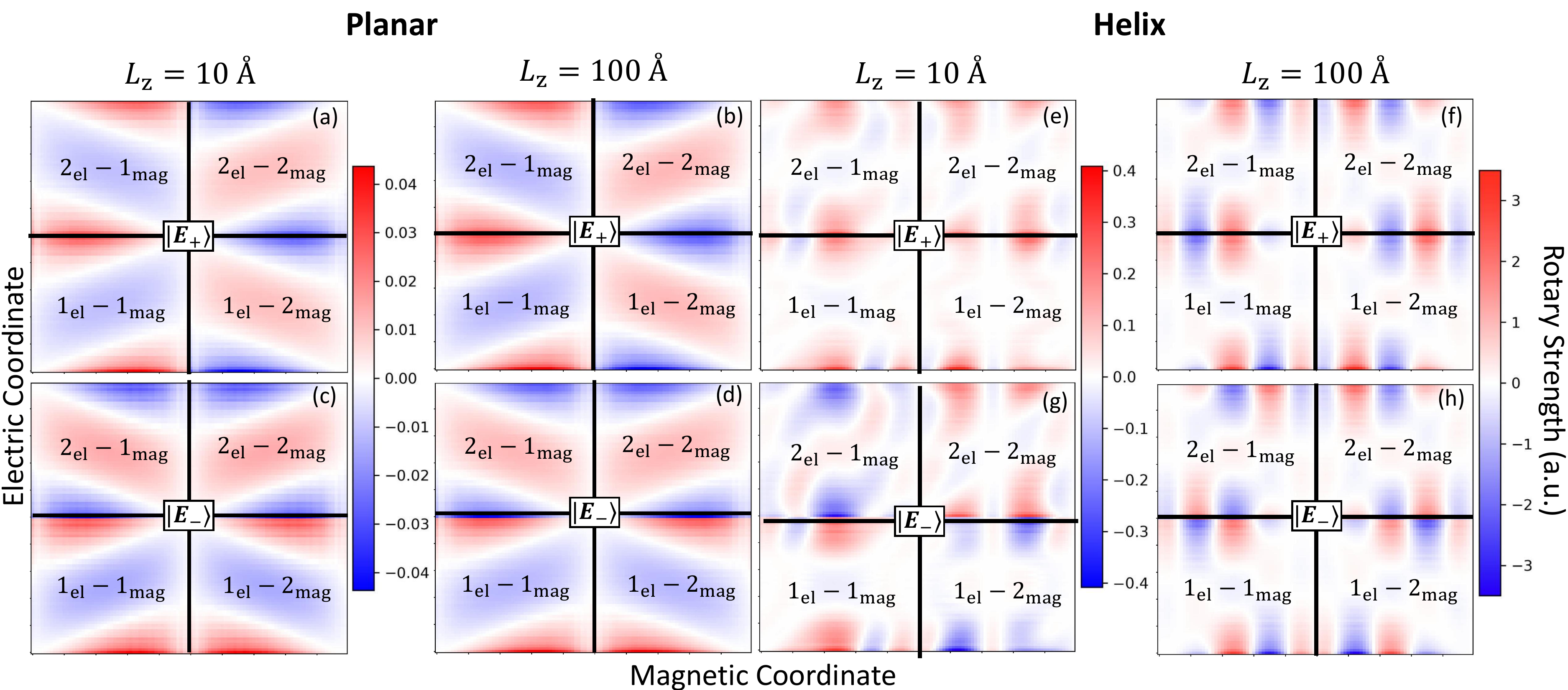}
    \caption{\footnotesize
    Transition chiral tensors (TCTs) for (a-d) achiral and (e-h) chiral geometries for fixed intermolecular angle $\theta$ = $0\degree$ at two intermolecular separations (a,c,e,g) $L_z = 10\AA$ (strongly coupled) and (b,d,f,h) $100 \mathrm{\AA}$ (weakly coupled) for both the (a,b,e,f) $|E_+\rangle$ and (c,d,g,h) $|E_-\rangle$ excitons. The atomic orbital contributions are summed into each chromophores' minimal C$_2$H$_2$ fragment as $R_{AB} = \sum_{\mu \in A} \sum_{\nu \in B} R_{\mu\nu}$ where $\{\mu,\nu\}$ are atomic orbitals and $\{A,B\}$ are the C$_2$H$_2$ fragments.
    }
    \label{FIG:chiral_achiral_dimer_TCTs_Lz_10_100_0DEG}
\end{figure*}

Due to the fact that each individual chromophore is achiral, the intramolecular components (\textit{i.e.} the diagonal blocks) are small compared to the intensities of the intermolecular components. For the planar chromophore, the transition electric dipole density and the transition magnetic dipole density are of opposite symmetry, which we have pointed out in a previous work.\cite{forde_helically_2026}. This implies the integration of their products (\textit{i.e.}, an even function multiplied by odd function yields an odd function whose integral is zero) will be negligible, which is observed in the diagonal blocks of the TCT's. However, when considering the case of coupled chromophores at a finite interaction angle $\theta \ne 0\degree$, there is interaction between transition electric dipole density and the transition magnetic dipole density on each chromophore, observed as the off-diagonal blocks. From the model perspective, this interaction arises in the gauge term in Eq.~\ref{EQ:R_DIMER}. The character and symmetry of the diagonal blocks will be discussed later in Figure~\ref{FIG:chiral_achiral_dimer_TCTs_Lz_10_100_0DEG}. The off-diagonal blocks showcase a clear symmetry, dependent on the choice of enantiomer (\textit{i.e.}, the angle $\theta$) and the constructive $|E_+\rangle$ (Figure~\ref{FIG:achiral_dimer_TCTs_Lz_100_PROJECTIONS}a,b) or deconstructive $|E_-\rangle$ (Figure~\ref{FIG:achiral_dimer_TCTs_Lz_100_PROJECTIONS}c,d) superposition of chromophore wavefunctions. For the case of $\theta = -50\degree$, the deconstructive $|E_-\rangle$ exciton exhibits a negatively signed response while the upper state constructive $|E_+\rangle$ yields a positive response. The reverse is true for $\theta = 50\degree$. Importantly, both off-diagonal blocks exhibit the same sign, leading to a cooperative interaction (or forward-backward symmetry) between chromophores.

In addition to the sign and symmetry, the fine structure of the off-diagonal blocks is also worth discussing, which is shared by all off-diagonal blocks from both angles and both hybrid excitons. For the 1$_\mathrm{el}$-2$_\mathrm{mag}$ component, the response is localized to the top and bottom edges with negative sign. The distribution extends across chromophore 2's magnetic contribution and exhibits a decreasing response closer to the center of chromophore 1's electric contribution. Summing chromophore 2's magnetic contribution would yield a bimodel function of chromophore 1's electric contribution with maxima at the edges of chromophore 1's electric coordinate (see Figure~\ref{FIG:achiral_dimer_TCTs_Lz_100_PROJECTIONS}e-h, black curve).\cite{forde_helically_2026} In contrast, summing over chromophore 1's electric contribution would yield a function centered at the middle of the chromophore 2's magnetic coordinate (see Figure~\ref{FIG:achiral_dimer_TCTs_Lz_100_PROJECTIONS}e-h,red curve). These one-dimensional projections indicate the spatial localization of the electric or magnetic contribution to the rotary strength, which has been discussed in a previous works.\cite{weight_influence_2026} Note that this projection analysis is only possible when working with the non-symmetrized TCT.\cite{weight_visual_2026}  This resembles the electric distribution when summing over the magnetic coordinate of a chromophore. Overall, the distribution of the off-diagonal elements suggests that the electric transition dipole character from one chromophores edges primarily interacts with the center of the other chromophore's magnetic transition dipole character.

Figure~\ref{FIG:achiral_E_D_R_ABS_CD_10_100}f-j presents similar data as shown in Figure~\ref{FIG:achiral_E_D_R_ABS_CD_10_100}a-e but for a fixed intermolecular separation of $L_z = 10~\mathrm{\AA}$, which represents the strongly coupled case. Aside from the exciton coupling deviating from the $\cos(\theta)$ ideal function (as discussed in Figure~\ref{FIG:geom_scheme}), the trends in the dipole strength and rotary strength are unaffected, except for the reduction in the intensity of the rotary strength by 10$\times$ due to the linear scaling of the rotary strength with the intermolecular separation shown in Eq.~\ref{EQ:R_DIMER} in \textbf{Methods}. Furthermore, the TCTs for the strongly coupled case ($L_z = 10~\mathrm{\AA}$) are shown in Supporting Figure~\ref{FIG:achiral_dimer_TCTs_Lz_10} and are very similar to the TCTs for the weakly coupled case (Figure~\ref{FIG:achiral_dimer_TCTs_Lz_100_PROJECTIONS}a-d).

\begin{figure}[t!]
    \centering
    \includegraphics[width=0.96\linewidth]{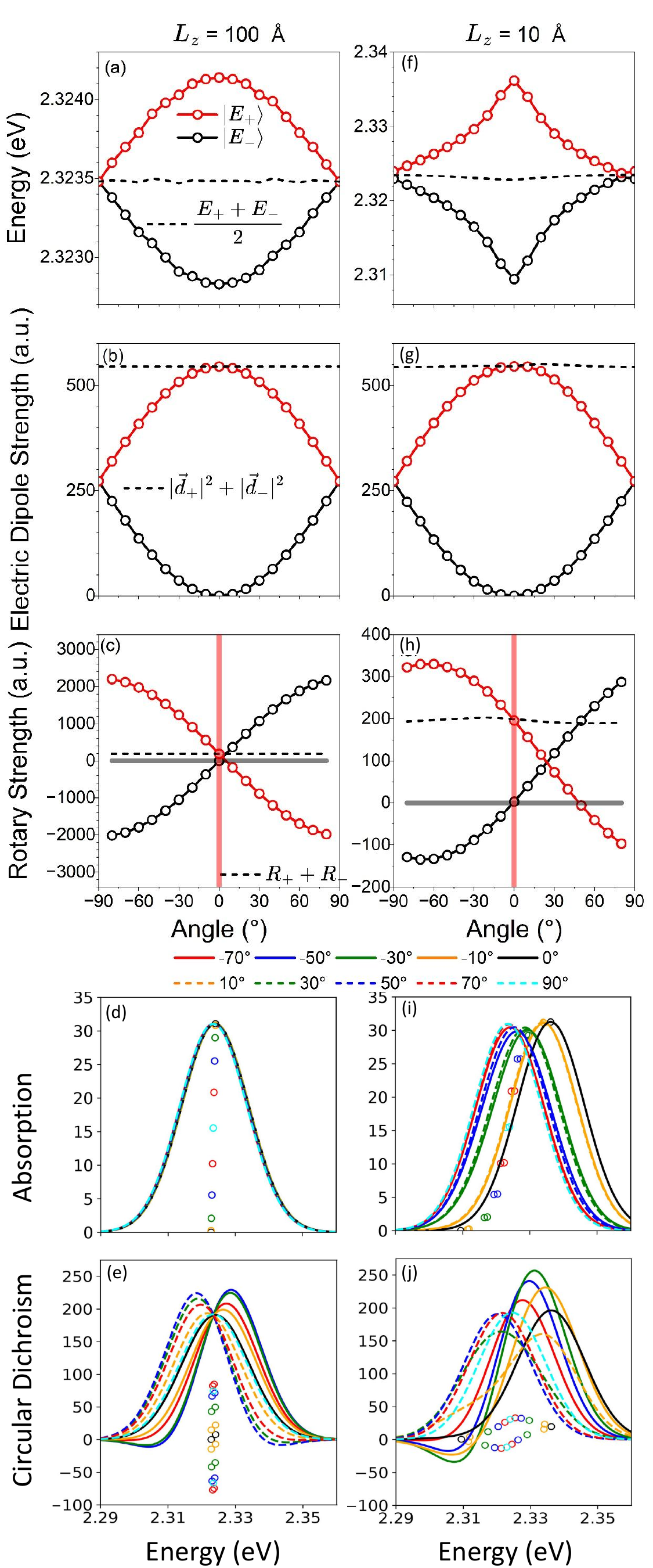}
    \caption{\footnotesize
    Exciton (a,f) energies $E_\pm$, (b,g) dipole strengths $|\vec{d}_\pm|^2$, (c,h) rotary strengths $R_\pm$, (d,i) absorption spectra and (e,j) CD spectra for a pair of \textit{chiral} chromophores separated by (a-d) $L_z = 100~\mathrm{\AA}$ (weakly coupled) and (e-h) $L_z = 10~\mathrm{\AA}$ (strongly coupled). The horizontal dashed black lines are conserved quantities for all coupling parameters. The spectra was generated with phenomenological Gaussian broadening $\sigma = 10$ meV. The symbols in the spectra indicate scaled oscillator and rotary strength values.
    }
    \label{FIG:chiral_E_D_R_ABS_CD_10_100}
\end{figure}

\begin{figure*}[t!]
    \centering
    \includegraphics[width=\linewidth]{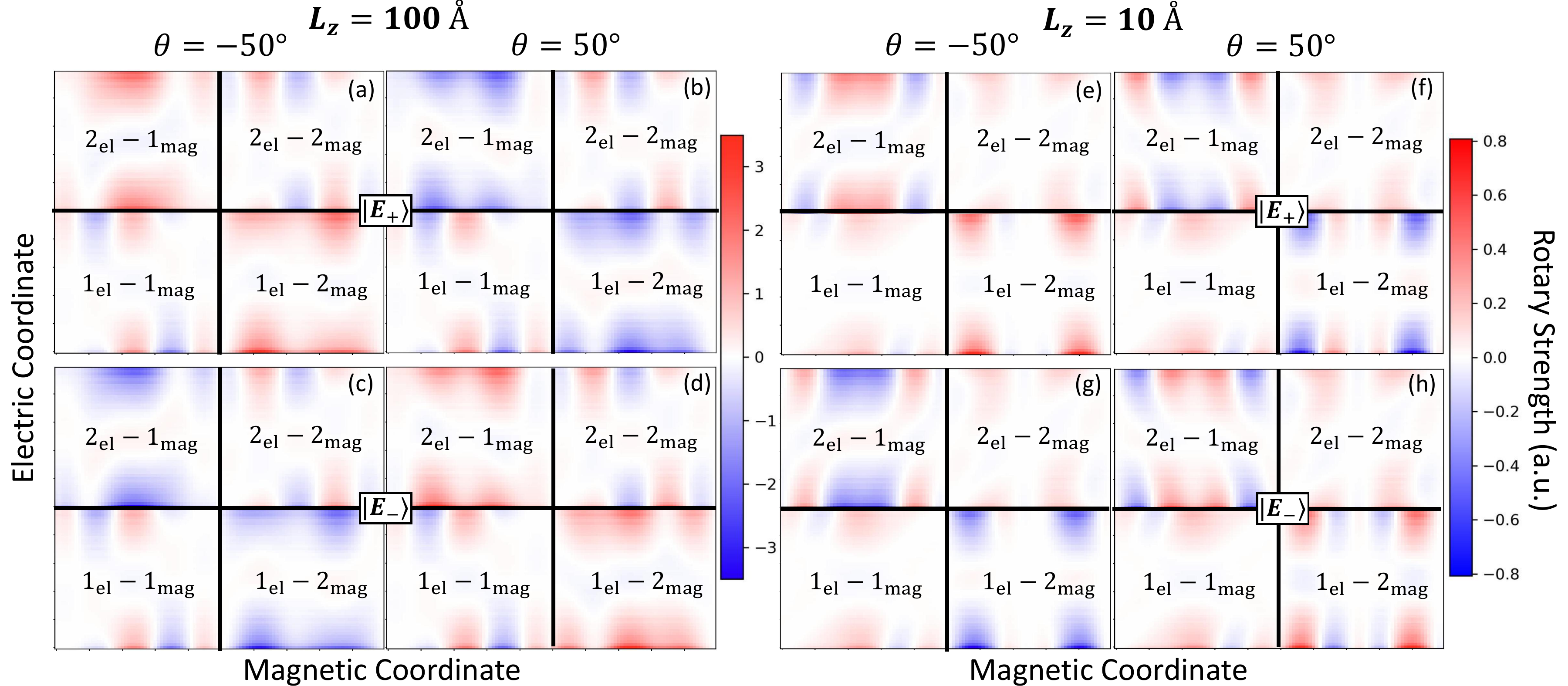}
    \caption{\footnotesize
    Transition chiral tensors (TCTs) for enantiomeric angles $\theta$ = (a,c,e,g) $-50\degree$ and (b,d,f,h) $50\degree$ for a dimer composed of two \textit{chiral} chromophores spaced at (a-d) $L_{z} = 100~\mathrm{\AA}$ (weakly coupled) and (e-h) $L_{z} = 10~\mathrm{\AA}$ (strongly, coupled) and for both the (a,b,e,f) $|E_+\rangle$ and (c,d,g,h) $|E_-\rangle$ excitons. The atomic orbital contributions are summed into each chromophores' minimal C$_2$H$_2$ fragment as $R_{AB} = \sum_{\mu \in A} \sum_{\nu \in B} R_{\mu\nu}$ where $\{\mu,\nu\}$ are atomic orbitals and $\{A,B\}$ are the C$_2$H$_2$ fragments.
    }
    \label{FIG:chiral_dimer_TCT_Lz_100_10}
\end{figure*}

{\bf Chiral Substituents}

Now we move to a similar discussion but with a set of chiral chromophores (as shown in Figure~\ref{FIG:geom_scheme}a,b, ``Helix''). The same planar polyacetylene chain was rotated such that it formed a helix which has exactly two rotations (pitch angle is $\approx14.69\degree$), described in more detail in Ref.~\citenum{forde_helically_2026}. The energy splitting (Figure~\ref{FIG:chiral_E_D_R_ABS_CD_10_100}a,f) and dipole strength  (Figure~\ref{FIG:chiral_E_D_R_ABS_CD_10_100}b,g) are nearly identical to that of the planar dimer system (Figure~\ref{FIG:achiral_E_D_R_ABS_CD_10_100}). However, the rotary strength shows interesting differences. 

First, in the strong coupling case (Figure~\ref{FIG:chiral_E_D_R_ABS_CD_10_100}g, $L_z = 10~\mathrm{\AA}$), the sinusoidal curves have now been shifted up and to the right by $\Delta R \approx 100$ a.u. and $\Delta \theta \approx 25\degree$, respectively. The shift up in rotary strength for both the $|E_+\rangle$ and $E_-\rangle$ states is due to to the intrinsic chirality of the helix chromophores. Note that the sum of both rotary strengths is now $R_+ + R_- \approx 200$ a.u., which is the sum of intrinsic chirality of the total system for any given angle. Recall that for the dimers composed of achiral chromophores, the sum of rotary strengths was always zero. The shift to the right by $\Delta \theta \approx 25\degree$ is simply an effect of the helical structures interacting at close range and thus has an asymmetry between positive and negative angles. The asymmetry in $\theta$ stems from the non-point-dipole-like interactions which helix segments between chromophores.

The absorption spectra in Figure~\ref{FIG:chiral_E_D_R_ABS_CD_10_100}d,i in exhibits nearly identical structure and the achiral chromophores in Figure~\ref{FIG:achiral_E_D_R_ABS_CD_10_100}d,i. The circular dichroism spectra, however, show interesting differences compared to the achiral chromophores case. In both weak and strong coupling cases (Figure~\ref{FIG:chiral_E_D_R_ABS_CD_10_100}e,j), a Cotton effect is still observed. However, the intensity of the negative feature is drastically reduced due to the shift up in rotary strength for the two hybrid exciton states. This evidences the competition between intrinsic and topological chirality. For the strongly coupled case (Figure~\ref{FIG:chiral_E_D_R_ABS_CD_10_100}h), the asymmetry in the interaction angle $\theta$ is prominent in that the negative feature disappears completely for the positive angle geometries, $\theta > 0$. Thus, evidence of coupled excitons can be distinguished in both achiral and chiral substituents and in both strongly and weakly coupled regimes. \textit{Importantly, we find that intrinsically chiral chromophores disrupt the Cotton effect due to the competition between intrinsic chirality and topological chirality.}

We next examine the TCTs for chiral chromophore cases in Figure~\ref{FIG:chiral_dimer_TCT_Lz_100_10}. As expected, the diagonal blocks of the TCTs are now activated due to the intramolecular chiroptical response. Notably, for the weakly coupled case ($L_z = 100~\mathrm{\AA}$), the off-diagonal blocks are dominated by a single sign of chiroptical response. For example, the 2-1 and 1-2 blocks of the $\theta = -50\degree$ $|E_+\rangle$ case is positive while the 2-1 and 1-2 blocks of the $\theta = -50\degree$ cases is negative, and \textit{vice versa} for the $|E_-\rangle$ state. However, in the strongly coupled case, due to the atom-to-atom interactions rather than point-dipole-like interactions, the off-diagonal blocks now exhibit interference effects (\textit{i.e.}, the appearance of both positive and negative features) between the electric and magnetic transition dipole interactions. The opposite is true for the diagonal blocks. For the weakly interacting case, the diagonal block are dominated by a single sign while the off-diagonal blocks now exhibit deconstructive interference. The strongly coupled case also exhibits this flip from off-diagonal-to-diagonal destructive-to-constructive response.

{\bf Symmetries of Achiral Intermolecular Configurations}

Finally, we examine the TCTs for the $\theta = 0\degree$ cases in both the weak and strong coupling regimes and for both the achiral (planar) and chiral (helix) chromophores in Figure~\ref{FIG:chiral_achiral_dimer_TCTs_Lz_10_100_0DEG}. The planar dimers (Figure~\ref{FIG:chiral_achiral_dimer_TCTs_Lz_10_100_0DEG}a-d) have no chiroptical response, as shown by the zero rotary strength for each hybrid excitonic state in Figure~\ref{FIG:achiral_E_D_R_ABS_CD_10_100}c,g. As such, the sum of the four TCTs blocks must sum to zero as $R^\pm = \sum_{A,B} R^\pm_{A,B} = 0$ where $\{A,B\}$ are the chromophore labels. Both the strongly interacting and weakly interacting regimes provide the same TCTs with the same symmetry but differ in their magnitudes, as discussed above due to their separation length. For the cases of the $|E_+\rangle$ transition, the left column (\textit{i.e.}, the 1-1 and 2-1 blocks) sum to a negative value while the right column (\textit{i.e.}, the 2-2 and 1-2 blocks) sum to the same magnitude but of positive sign, thus canceling the overall summation of all four blocks. Thus, one could say that the cross-chromophore blocks, 1-1 to 2-2 or 1-2 to 2-1, provide equal and opposite local chiral responses. This implies that there is chirality in their atom-to-atom interactions, but these local contributions are exactly canceled by the inversion symmetry of the system.

The chiral chromophore case (Figure~\ref{FIG:chiral_achiral_dimer_TCTs_Lz_10_100_0DEG}e-h) has similar interpretations but are augmented by the fact that the individual chromophores retain some chiroptical response. For example in the strongly coupled case (Figure~\ref{FIG:chiral_achiral_dimer_TCTs_Lz_10_100_0DEG}e,g), the lower hybrid exciton $|E_-\rangle$ has no chiroptical response (see Figure~\ref{FIG:chiral_E_D_R_ABS_CD_10_100}c, black) at $\theta = 0\degree$ while the upper state $|E_+\rangle$ has an appreciable response. In this case, the blocks of the TCT for the $|E_+\rangle$ state are positive (with minimal deconstructive inference in the off-diagonal blocks) while blocks of the $|E_-\rangle$'s TCT are mixed, providing an overall negligible response. The weakly coupled case (Figure~\ref{FIG:chiral_achiral_dimer_TCTs_Lz_10_100_0DEG}f,h) is slightly more complicated since both states seems to contain substantial interference. However, looking at the block-summed (\textit{i.e.}, summing all atomic orbitals for each chromophore) 4$\times$4 TCT in Supporting Figure~\ref{SUPP_FIG:TCT_4x4_0DEG}f,h, the lower state's $|E_-\rangle$ diagonal blocks sum to a positive value while the off-diagonal blocks perfectly cancel by summing to the same value but with a negative sign. For the upper state $|E_+\rangle$, however, all blocks sum to a positive value.

\section{Conclusions}

In this work, we have presented a systematic computational investigation into the microscopic origins of circular dichroism in coupled molecular dimers supporting delocalized excitonic excitations. Using polyacetylene chromophores as a model system to disentangle the interplay between intermolecular (topological) chirality and intramolecular (intrinsic) chirality. By combining TDDFT electronic structure simulations with a recently developed transition chiral tensor (TCT) analysis, we have provided both a quantitative and visually intuitive framework for understanding how exciton coupling strength, intermolecular separation $L_z$, and relative electric transition dipole orientation $\theta$ collectively shape the chiroptical response of interacting chromophores.

Our results demonstrate that even pairs of individually achiral chromophores can exhibit a robust, coupling-induced chiroptical response purely as a consequence of their relative geometric arrangement (a topological chirality), with the rotary strength vanishing at both maximized coupling ($\theta = 0\degree$) and maximizing at vanishing coupling ($\theta = \pm90\degree$). This behavior directly reflects a fundamental trade-off between exciton delocalization (and thus absorptive intensity) and the symmetry required to generate a nonzero chiral signal. Spectrally, this effect manifests only at intermediate angles, in which case a Cotton effect appears, a standard interpretation of derivative features in CD spectra. Importantly, for achiral and chiral chromophores, the CD spectra exhibits configuration-dependent spectral signatures while the absorption spectra does not.

When intrinsic chromophore chirality is introduced (via helical chromophores), we find that the intrinsic and topological chiral responses superpose in a strongly angle- and distance-dependent manner. This interplay produces asymmetric rotary strength and spectral profiles, reduced (or hidden) Cotton features, and even the complete suppression of one of the two chiral signatures at large coupling strengths (small intermolecular separation). This highlights that the observable CD spectrum of a coupled chromophore system is not merely a simple sum of its individual chromophore responses but instead emerges from a delicate and highly tunable balance between local and global chirality.

The transition chiral tensor analysis further reveals that this balance is encoded in the spatial distribution of electric and magnetic transition dipole contributions both within and between chromophores. Systems composed of achiral chromophores produce clean, single-sign intra- and inter-molecular blocks, whereas chiral chromophore systems exhibit pronounced interference patterns. Critically, we show that the diagonal (intramolecular) and off-diagonal (intermolecular) contributions to the rotary strength trade dominance depending on both the coupling regime and the intrinsic chirality of the constituent chromophores, offering a clear microscopic picture of how chiral signals build up, cancel, or reinforce one another across a coupled molecular system.

Taken together, these findings suggest practical design principles for engineering chiroptical materials: the sign, magnitude, and spectral signature of CD response can be systematically tuned through control of intermolecular distance and orientation, independent of (or in cooperation with) the intrinsic chirality of the constituent units. This work exemplifies interpretation of exciton-coupled CD spectra in molecular aggregates, supramolecular assemblies, and chiral nanostructures. The TCT methodology introduced here offers a general and transferable tool for visualizing and rationalizing chiroptical responses in more complex multi-chromophore systems, including those relevant to chiral-induced spin selectivity and other emerging quantum and photonic technologies.

\section{Methods}
{\bf Transition Chiral Tensor}

Here, we provide details on the local decomposition of the transition dipoles and the Transition Chiral Tensor analysis. In the collective electronic oscillator (CEO) picture\cite{tretiak_density_2002,tretiak_recursive_1996,tretiak_collective_1996,weight_visual_2026}, the electronic coordinate (symmetric) and momentum (anti-symmetric) are written as
\begin{equation}
    \textbf{Q}^{0\alpha} = \frac{1}{\sqrt{2}}\big( \boldsymbol{\xi}^{0\alpha} + {\boldsymbol{\xi}^{0\alpha}}^\dagger\big),
\end{equation}

and
\begin{equation}
    \textbf{P}^{0\alpha} = \frac{1}{\sqrt{2}}\big( \boldsymbol{\xi}^{0\alpha} - {\boldsymbol{\xi}^{0\alpha}}^\dagger\big),
\end{equation}
respectively.

The rotary strength between the ground and excited RPA state can again be decomposed into electric and magnetic components as
\begin{align}\label{EQ:EXCITONIC_EL_DIPOLE}
    \vec{d}^{0\alpha} &= \langle 0 | \vec{\hat{r}} | \alpha \rangle = \mathrm{Tr}[\vec{\boldsymbol{d}}~\textbf{Q}^{0\alpha}]\nonumber\\
    &= \sum_{ia} \vec{d}_{ia} Q^{0\alpha}_{ia},
\end{align}
and
\begin{align}\label{EQ:EXCITONIC_MAG_DIPOLE}
    \vec{m}^{0\alpha} &= \langle 0 | \vec{\hat{m}} | \alpha \rangle= \mathrm{Tr}[\vec{\boldsymbol{m}}~\textbf{P}^{0\alpha}]\nonumber\\
    &= \sum_{ia} \vec{m}_{ia} P^{0\alpha}_{ia},
\end{align}
respectively, with occupied orbital $i$ and virtual orbital $a$. The MO integrals for the electric and magnetic dipole moments are obtained by contraction over the atomic orbital (AO) basis, $\{\mu,\nu\}$. The AO integrals $\vec{d}_{\mu\nu} = \langle \mu | \vec{\hat{d}} | \nu\rangle$ and $\vec{m}_{\mu\nu} = \langle \mu | \vec{\hat{m}} | \nu\rangle$, MO expansion coefficients $c_{\mu i}$, and the transition densities $\xi^{0\alpha}_{ia}$ for each transition $0\rightarrow\alpha$ are directly obtained by the Gaussian16 software package.\cite{frisch_gaussian16_2016} Here, $\hat{d} = -\sum_{i} \vec{\hat{r}}_i + \sum_I Z_I \vec{R}_I \mathbb{I}$ is the electric dipole operator, summing over each electron $i$ and nucleus $I$, and $\hat{m} = \sum_{i} \vec{\hat{r}}_i \times \vec{\hat{\nabla}}_i$ is the magnetic dipole operator with $\vec{\hat{\nabla}} = i\vec{\hat{p}}$ is the usual gradient operator. The corresponding rotary strength in the excitonic basis $\{\alpha\}$ is then written as
\begin{equation}
    R^{0\alpha} = \vec{d}^{0\alpha} \cdot \vec{m}^{\alpha 0}.
\end{equation}

The rotary strength in the excitonic basis is then visualized according to Ref.~\citenum{weight_visual_2026}. The transition chiral tensor is constructed as
\begin{equation}
    R_{\mu\nu} = \vec{d}^{0\alpha}_{\mu\mu} \cdot \vec{m}^{0\alpha}_{\nu\nu},
\end{equation}
where $\vec{d}^{0\alpha}_{\mu\mu} = \langle \mu | \boldsymbol{\vec{d}}~\boldsymbol{Q}^{0\alpha}|\mu\rangle$ and $\vec{\boldsymbol{m}}^{0\alpha}_{\mu\mu} = \langle \mu | \vec{\boldsymbol{m}}~\boldsymbol{P}^{0\alpha}|\mu\rangle$ are the components of the trace in Eqs.~\ref{EQ:EXCITONIC_EL_DIPOLE} and~\ref{EQ:EXCITONIC_MAG_DIPOLE} in the AO basis. 
This matrix can then be symmetrized as $(R^{0\alpha}_{\mu\nu} + R^{0\alpha}_{\mu\nu})/2 \rightarrow R^{0\alpha}_{\mu\nu}$. Note that the asymmetric part, $(R^{0\alpha}_{\mu\nu} - R^{0\alpha}_{\mu\nu})/2$, does not contribute to the total rotary strength but can, and does, contribute to any visualization. The symmetrization of the TCT mixes the electric and magnetic components such that there are no longer two independent coordinates. This removes the ability to compare directionality/origin of the interaction, \textit{e.g.}, the magnetic dipole of chromophore 1 interacts with the electric dipole of chromophore 2. Therefore, we choose not to symmetrize the TCTs in this work to retain the electric and magnetic components of the interactions as independent axes, which are destroyed upon symmetrization.

The $\boldsymbol{Q}^{0\alpha}$ and $\boldsymbol{P}^{0\alpha}$ can be transformed into the AO basis as
\begin{equation}\label{EQ:Q_mo_to_ao}
    Q_{\mu\nu}^{0\alpha} = \sum_{ia} c_{\mu i}~Q^{0\alpha}_{ia}~c_{\nu a}
\end{equation}
and
\begin{equation}\label{EQ:P_mo_to_ao}
    P_{\mu\nu}^{0\alpha} = \sum_{ia} c_{\mu i}~P^{0\alpha}_{ia}~c_{\nu a}.
\end{equation}
Finally, the transition chiral tensors are atom-localized by summing over the AOs localized to each atom or fragment $\{A,B\}$ as 
\begin{equation}\label{EQ:CONDENSED_R_TENSOR}
    R^{0\alpha}_{AB} = \sum_{\mu \in A} \sum_{\mu \in B} R^{0\alpha}_{\mu\nu}.
\end{equation}
Note that the rotatory strength is conserved
\begin{equation}\label{EQ:R_from_Rmn_RAB}
    R^{0\alpha} = \sum_{\mu,\nu}R_{\mu\nu}^{0\alpha} = \sum_{A,B} R_{AB}^{0\alpha}.
\end{equation}

{\bf Model Hamiltonian}

Given two chromophores, labeled as ``1'' and ``2'', the interaction Hamiltonian can be written as
\begin{equation}\label{EQ:EXCITON_HAM}
    \hat{H} = \begin{bmatrix}
        \omega & J_{12}\\
        J_{12} & \omega
    \end{bmatrix}
\end{equation}
where $J_{12}$ is the coupling matrix elements and $\omega$ is the bare chromophore transition energy. In this work, we assume all chromophores to be identical. Although, all results presented here can be extended to non-identical chromophores.

In principle, the excitonic coupling involving both electric and magnetic couplings, which can be truncated at the point-dipole level in the multi-polar expansion. These interactions can be written as,
\begin{align}\label{EQ:J_EL}
    J^\mathrm{el}_{12} &=\frac{ \vec{d}_1 \cdot \vec{d}_2 - 3 (\vec{d}_1 \cdot \hat{R}_{12}) (\vec{d}_2 \cdot \hat{R}_{12}) }{|\vec{R}_{12}|^3}\nonumber\\
    &= |\vec{d}_1| |\vec{d}_2| \cos(\theta)
\end{align}
for the electric transition dipole and as
\begin{align}
    J^\mathrm{mag}_{12} &= \frac{1}{c^2}\frac{ \vec{m}_1 \cdot \vec{m}_2 - 3 (\vec{m}_1 \cdot \hat{R}_{12}) (\vec{m}_2 \cdot \hat{R}_{12}) }{|\vec{R}_{12}|^3}\nonumber\\
    &= |\vec{m}_1| |\vec{m}_2| \cos(\theta)
\end{align}
for the magnetic transition dipole. Here, $\vec{d}_i = \langle \mathrm{S}^{(i)}_1|\vec{\hat{d}}| \mathrm{G}\rangle$ is the ground-to-excited (S$_0$-to-S$_1$) electric transition dipole moment for chromophore $i$, $\vec{m}_i = \langle \mathrm{S}^{(i)}_1|\vec{\hat{m}}| \mathrm{G}\rangle$ is the magnetic transition moment, and $\vec{R}_{12} = \vec{R}_1 - \vec{R}_2$ is the intermolecular position vector connecting the center-of-mass positions of each chromophore. In both cases, the second term is zero since we keep all geometries in this work vertically shifted (along the $z$-axis) and simply rotated about the $z$-axis by angle $\theta$. We note that the coupling element for the magnetic dipole is usually neglected since it contains an additional factor of $1/c^2$ where $c \approx 137$ is the speed of light. Note that $c^2 = \frac{1}{\epsilon_0 \mu_0}$ and $\epsilon_0 = \frac{1}{4\pi}$. We will also drop this term in the model Hamiltonian and assume that the energy splitting is dominated by the electric dipole contribution, \textit{i.e.}, $J_{12} = J^\mathrm{el}_{12}$. However, importantly, we keep the magnetic transition dipole for the spectroscopy, as it is equally as important as the electric transition dipole in the calculation of the circular dichroism spectroscopy.

The transition electric (magnetic) dipole is defined as $\vec{d}^{\alpha\beta}_{i} =  \langle \psi_\alpha | \vec{\hat{d}} | \psi_\beta \rangle$ ($\vec{m}^{\alpha\beta}_{i} =  \langle \psi_\alpha | \vec{\hat{m}} | \psi_\beta \rangle$) where $|\psi_\alpha\rangle$ is the $\alpha_\mathrm{th}$ electronic state. $\vec{\hat{d}} = -\sum_{i} \vec{\hat{r}}_i$ is the electronic part of the electric dipole operator and $\vec{\hat{m}} = -\sum_{i} \vec{\hat{r}}_i \times \vec{\hat{\nabla}}_i$, with $\vec{\hat{\nabla}} = -i\vec{\hat{p}}$, is the electronic part of the magnetic dipole operator.

Upon diagonalization of Eq.~\ref{EQ:EXCITON_HAM}, $\hat{H} |E_{\pm}\rangle = E_{\pm}|E_{\pm}\rangle$, the hybridized exciton states, $|E_{\pm}\rangle = C_1|1\rangle + C_2|2\rangle$, and energies, $E_{\pm} = \omega \pm J_{12}$, can be used to calculate the model absorption and CD spectra. The collective electric dipole operator is written as
\begin{align}
    \vec{d}_\mathrm{\pm} &= \langle E_{\pm} | \vec{\hat{d}} | \mathrm{G} \rangle = \sum_{i} C_i \langle i | \vec{\hat{d}} | \mathrm{G} \rangle\nonumber\\
    &= \sum_{i} C_i~\vec{d}_i
\end{align}
whereas the magnetic dipole picks up an additional term to correct for origin invariance (a gauge term) as
\begin{align}
    \vec{m}_\mathrm{\pm} &= \langle E_{\pm} | \vec{\hat{M}} +  \vec{\hat{m}} | \mathrm{G} \rangle\nonumber\\
    &= \sum_{i} C_{i} \langle i | \vec{\hat{M}} +  \vec{\hat{m}} | \mathrm{G} \rangle\nonumber\\
    &= \sum_{i} C_{i} \langle i | \vec{\hat{M}}| \mathrm{G}  \rangle + C_i \langle i | \vec{\hat{m}}| \mathrm{G} \rangle\nonumber\\
    &= \sum_{i} C_{i} (\vec{M}_i + \vec{m}_i)
\end{align}
where $\vec{\hat{M}} = -\vec{\hat{R}} \times \vec{\hat{\nabla}}$ with $\vec{\hat{R}}$ is the center-of-mass operator which acts as $\vec{\hat{R}} | 1 \rangle = \vec{R}_1 | 1 \rangle$.

\begin{align}
    \langle i | \vec{\hat{M}}| \mathrm{G} \rangle &= \langle i | \vec{\hat{R}} \times \vec{\hat{\nabla}} | \mathrm{G}  \rangle\nonumber\\
    &= \omega \langle i | \vec{\hat{R}} \times \vec{\hat{d}} | \mathrm{G}  \rangle\nonumber\\
    &= \omega \vec{R}_i \times \langle i | \vec{\hat{d}} | \mathrm{G} \rangle\nonumber\\
    &= \omega \vec{R}_i \times \vec{d}_i
\end{align}
Then, the rotary strength of the hybrid system can be calculated as
\begin{align}\label{EQ:R_DIMER}
    R_{\pm} &= \vec{d}_\mathrm{\pm} \cdot \vec{m}_\mathrm{\pm}\nonumber\\
    &= (\sum_{i} C_i \vec{d}_i) \cdot \big(\sum_{j} C_j (\vec{M}_j + \vec{m}_j)\big)\nonumber\\
    &= \sum_{ij} C_i C_j (\vec{d}_i \cdot \vec{M}_j +  \vec{d}_i \cdot \vec{m}_j)\nonumber\\
    &= \sum_{ij} C_i C_j (\vec{d}_i \cdot \omega \langle j | \vec{\hat{R}} \times \vec{\hat{d}} | \mathrm{G}  \rangle +  \vec{d}_i \cdot \vec{m}_j)\nonumber\\
    &= \sum_{ij} C_i C_j (\vec{d}_i \cdot \omega (\vec{R}_j \times \vec{d}_j) +  \vec{d}_i \cdot \vec{m}_j)\nonumber\\
    &= \sum_{ij} C_i C_j( \omega \vec{R}_j \cdot (\vec{d}_j \times \vec{d}_i) + \vec{d}_i \cdot \vec{m}_j)\nonumber\\
    &= \omega \sum_{ij} C_i C_j \vec{R}_j \cdot (\vec{d}_j \times \vec{d}_i) \nonumber\\
        &~~~~~~~~~~~~~~~+ \sum_{ij} C_i C_j \vec{d}_i \cdot \vec{m}_j\nonumber\\
    &= \omega \sum_{i < j} C_i C_j \vec{R}_{ij} \cdot (\vec{d}_i \times \vec{d}_j) \nonumber\\
        &~~~~~~~~~~~~~~~+ \sum_{ij} C_i C_j \vec{d}_i \cdot \vec{m}_j
\end{align}
Here, we have used the commutation relation $[\hat{H}, \hat{\hat{d}}]$ to obtain the conversion between position and momentum matrix elements, $\vec{p}_{ab} = iE_{ab}~\vec{d}_{ab}$ or $\vec{\nabla}_{ab} = E_{ab}~\vec{d}_{ab}$. We also used the following vector identities: $\vec{a} \times \vec{a} = \vec{0}$, $\vec{a} \cdot (\vec{b} \times \vec{c}) = \vec{b} \cdot (\vec{c} \times \vec{a})$, and $(\vec{a} \times \vec{b}) = - (\vec{b} \times \vec{a})$. Note that in some works, the term $\sum_{ij} C_i C_j \vec{d}_i \cdot \vec{m}_j$ is rewritten as $\frac{1}{2}\sum_{i j} (\vec{d}_i \pm \vec{d}_j) \cdot (\vec{m}_i \pm \vec{m}_j)$,\cite{berova_exciton_1999} which is only true for two identical chromophores.

\section{Computational Details}
Polyacetylene chain (H$_2$CC$_{n-2}$H$_{n-2}$CH$_2$, with $n = 100$) restricted to lie in the XY-plane and whose chromophore units form a perfect line (\textit{i.e.}, no curvature of the polymer in either x or y directions). The bond alternation is fixed and identical within each chromophore, taken from the center of an optimized structure. The optimized structure was not directly used since it was slightly curved, thus activating a chiral response, although small. For all \textit{ab initio} calculations, $\omega$B97XD/6-31G* long-range-corrected hybrid functional and basis set was used. For the TD-DFT simulations, the lowest 30 excited singlet states were used in all calculations. All electronic structure calculations were performed using the Gaussian16 software package.\cite{frisch_gaussian16_2016}

\begin{acknowledgement}
{\small
B.M.W. appreciates the support of the Director's Postdoctoral Fellowship at Los Alamos National Laboratory (LANL), which is funded by the Laboratory Directed Research and Development (LDRD) at LANL. LANL is operated by Triad National Security, LLC, for the National Nuclear Security Administration of the US Department of Energy (Contract No. 89233218CNA000001). This research is supported by the U.S. Department of Energy (DOE), Office of Science, Basic Energy Sciences, as part of the CHIME, a Microelectronics Science Research Center (MSRC). 
This work was performed in part at the Center for Integrated Nanotechnology (CINT) at LANL, a U.S. DOE and Office of Basic Energy Sciences user facility. This research used resources provided by the Darwin testbed, which is funded by the Computational Systems and Software Environments subprogram of LANL's Advanced Simulation and Computing program (NNSA/DOE), as and Institutional Computing (IC) program.
}
\end{acknowledgement}

{\small
\bibliography{main.bib}
}


\end{document}


\begin{figure}[t!]
    \centering
    \includegraphics[width=0.8\linewidth]{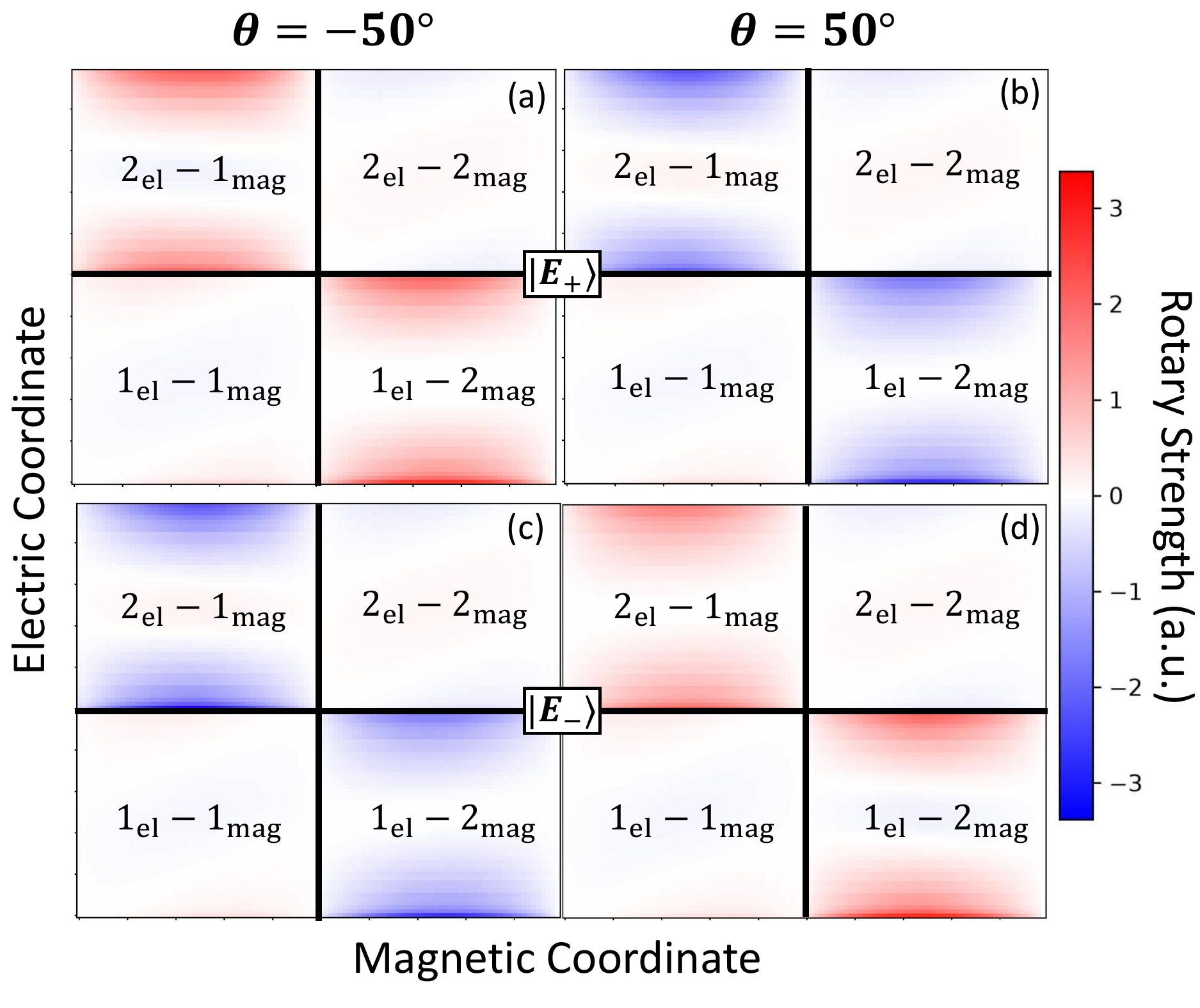}
    \caption{\footnotesize
    Transition chiral tensors (TCTs) for enantiomeric angles $\theta$ = (a,c,e,g) $-50\degree$ and (b,d,f,h) $50\degree$ for a dimer composed of two \textit{achiral} monomers spaced at (a-d) $L_{z} = 10~\mathrm{\AA}$ (strongly, coupled) and for both the (a,b) $|E_+\rangle$ and (c,d) $|E_-\rangle$ excitons. The atomic orbital contributions are summed into each monomers' minimal C$_2$H$_2$ fragment as $R_{AB} = \sum_{\mu \in A} \sum_{\nu \in B} R_{\mu\nu}$ where $\{\mu,\nu\}$ are atomic orbitals and $\{A,B\}$ are the C$_2$H$_2$ fragments.
    }
    \label{FIG:achiral_dimer_TCTs_Lz_10}
\end{figure}

\begin{figure}[t!]
    \centering
    \includegraphics[width=0.5\linewidth]{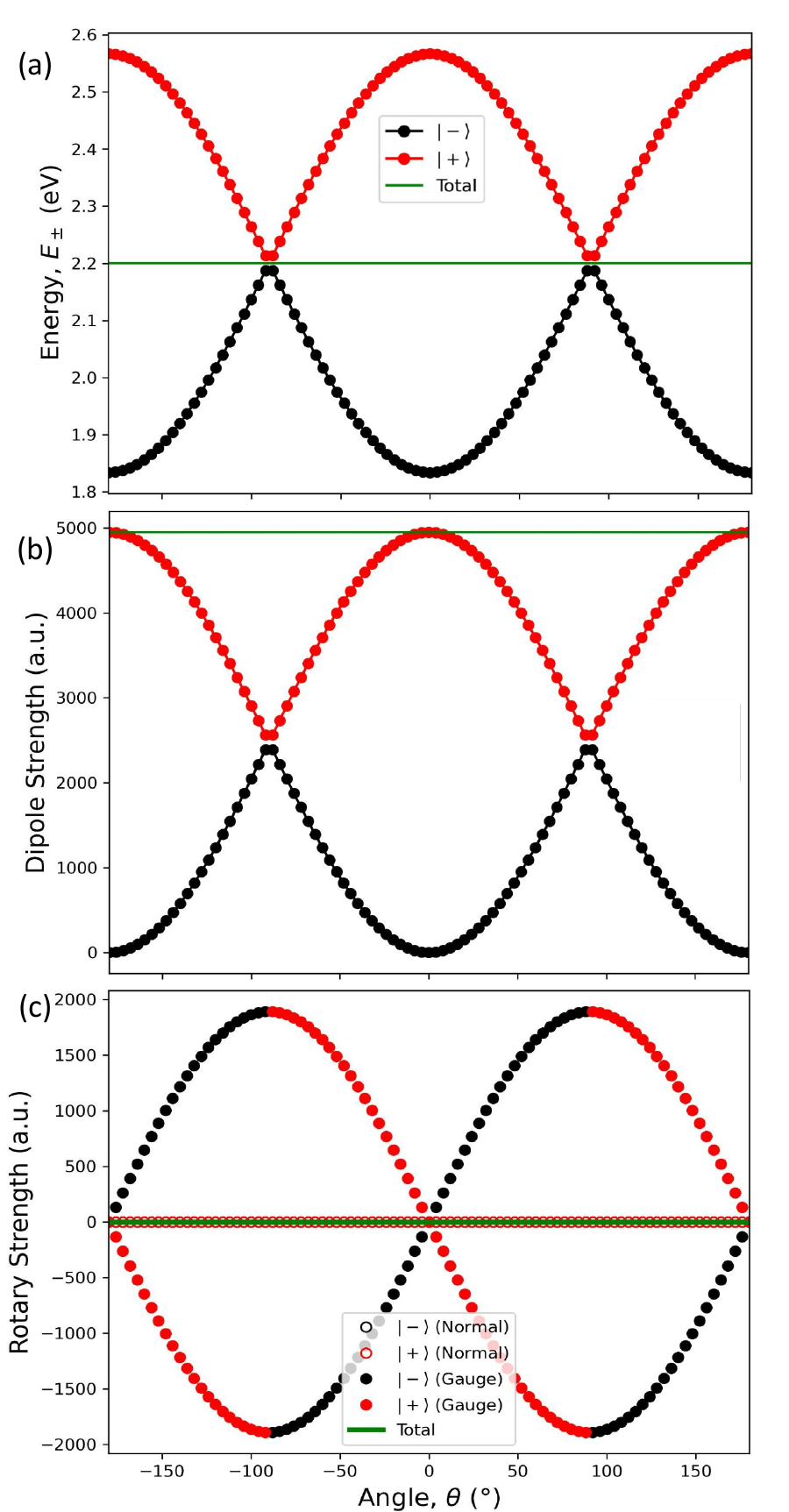}
    \caption{\footnotesize
    Model coupled exciton (a) energies $E_\pm$, (b) dipole strength $|\vec{d}_\pm|^2$, and (c) rotary strength $R_\pm$ based upon \textit{ab initio} data for the monomer (transition energy $\omega$, electric transition dipole $\vec{d}_{01}$, and magnetic transition dipole $\vec{m}_{01}$).
    }
    \label{FIG:model_achiral}
\end{figure}

\begin{figure}[t!]
    \centering
    \includegraphics[width=\linewidth]{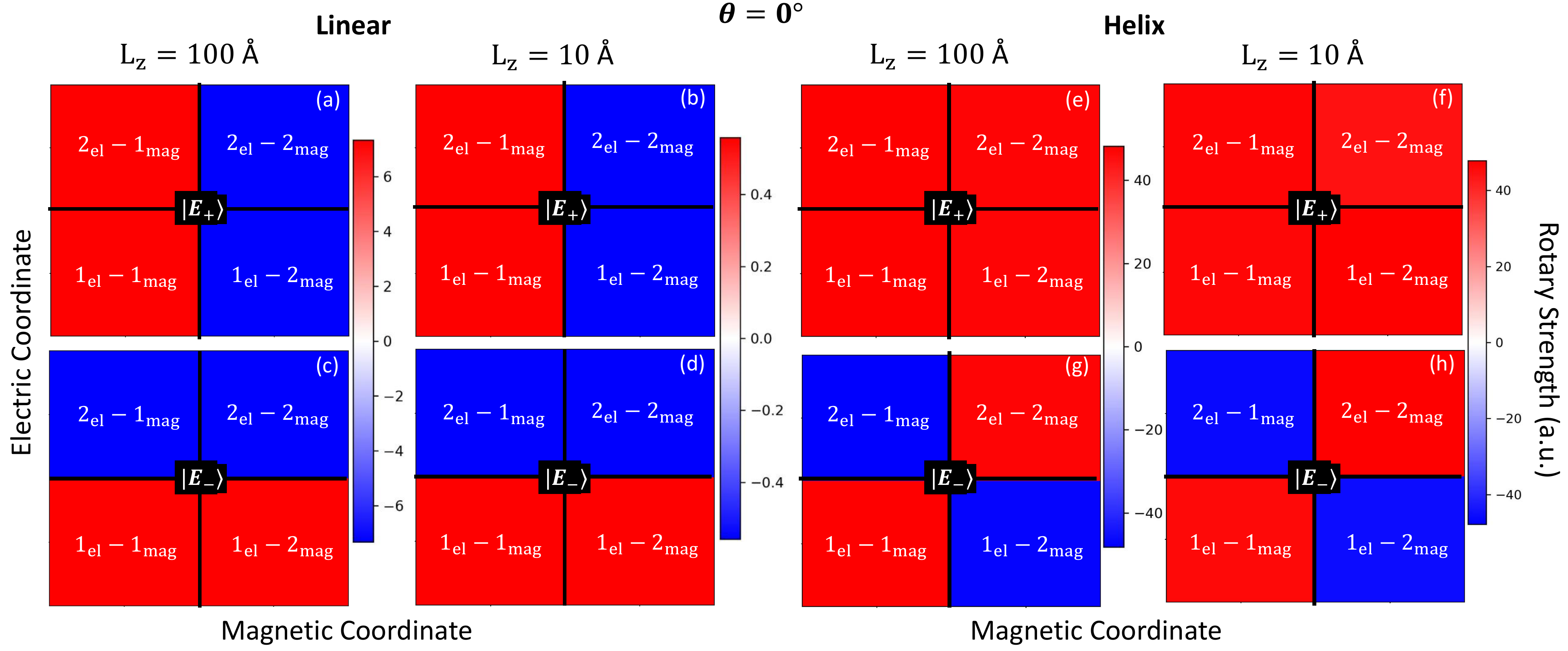}
    \caption{\footnotesize
    Transition chiral tensors (TCTs) for (a-d) achiral and (e-h) chiral geometries for fixed intermolecular angle $\theta$ = $0\degree$ at two intermolecular separations (a,c,e,g) $L_z = 10\AA$ (strongly coupled) and (b,d,f,h) $100 \mathrm{\AA}$ (weakly coupled) for both the (a,b,e,f) $|E_+\rangle$ and (c,d,g,h) $|E_-\rangle$ excitons. The atomic orbital contributions are summed into each monomers $R_{AB} = \sum_{\mu \in A} \sum_{\nu \in B} R_{\mu\nu}$ where $\{\mu,\nu\}$ are atomic orbitals and $\{A,B\}$ are the monomers.
    }
    \label{SUPP_FIG:TCT_4x4_0DEG}
\end{figure}